\documentclass[12pt]{article}
\input epsf
\newcommand{\sect}[1]{\section{#1}\setcounter{equation}{0}}

\renewcommand{\*}{ \hspace{-6pt}&=&\hspace{-6pt} }

\def\IR{{\rm I\!R}}

\def\Tr{{\rm Tr}}
\def\ZZ{{\bf Z}_2}
\begin{document}

\newpage
\bigskip
\hskip 3.7in\vbox{\baselineskip12pt
\hbox{DTP/00/17}
\hbox{hep-th/0002244}}

\bigskip
\bigskip
\bigskip
\bigskip

\centerline{\large \bf Orientifolds, M--Theory, and the $ABCD$'s of
the Enhan\c{c}on}
\bigskip\bigskip
\bigskip
\bigskip

\centerline{{\bf Laur
J\"arv\footnote{laur.jarv@durham.ac.uk}$^,$\footnote{Also:
Institute of
Theoretical Physics, University of Tartu, Estonia} and Clifford
V. Johnson\footnote{c.v.johnson@durham.ac.uk}}}

\bigskip
\bigskip
\bigskip
\centerline{Centre for Particle Theory}
\centerline{Department of Mathematical Sciences}
\centerline{University of Durham, Durham DH1 3LE}
\centerline{England, U.K.}

\bigskip
\medskip

\begin{abstract}
\baselineskip=16pt Supergravity solutions related to large $N$ $SU(N)$
pure gauge theories with eight supercharges have recently been shown
to give rise to an ``enhan\c con'', a new type of hypersurface made of
D--branes. We show that enhan\c cons also arise in similar situations
pertaining to $SO(2N+1)$, $USp(2N)$ and $SO(2N)$ gauge theories, using
orientifolds. Enhan\c cons therefore appear to come in types $A$, $B$,
$C$, and~$D$.  The latter three differ globally from type $A$ by
having an extra $\ZZ$ identification, and are distinguished locally by
their subleading behaviour in large $N$.  We focus mainly on $2{+}1$
dimensional gauge theory, where a relation to M--theory and the
Atiyah--Hitchin and Taub--NUT manifolds enables the construction of
the smooth supergravity solution and the study of some of the~$1/N$
corrections. The role of the enhan\c con in eleven dimensional
supergravity is also uncovered. There is a close relation to certain
multi--monopole moduli space problems.
\end{abstract}
\newpage
\baselineskip=18pt
\setcounter{footnote}{0}


\section{Opening Remarks}
In studying brane configurations related to large $N$ $SU(N)$ pure
gauge theories with eight supercharges, the authors of ref.\cite{jpp}
considered the BPS supergravity solutions which ought to result in
taking $gN$ large, where $g$ is the string coupling, and $N$ is the
number of constituent branes.  Such supergravity solutions are
afflicted by a naked singularity known as a
``repulson''\cite{repulsive} which is unphysical, and incompatible
with the physics of the gauge theory.

Upon closer examination (by investigating how such a geometry could
have arisen by constructing it out of a large number of its
constituent BPS parts) it was argued\cite{jpp} that the repulson is
not present. The supergravity solution may only be taken as physical
down to a radius of closest approach. At that locus of points (in the
case of a $(p+1)$--dimensional gauge theory, it is a $(4-p)$--sphere,
$S^{4-p}$), there is an enhanced gauge symmetry in the parent string
theory and new physics, consistent with the related $SU(N)$ gauge
theory, takes over.

That locus of points ---called the ``enhan\c con''--- is new type of
hypersurface essentially made of D--branes. The entire curved geometry
is produced by a large number of identical BPS objects. An individual
unit, when separated from all the others, is an object which has a
simple description in terms of (roughly) a pair of D--branes, one of
which is partially wrapped on a $K3$ surface, and the other which is
induced by the wrapping (see later). It it therefore a sharply
localised and heavy object.  Upon approaching the geometry produced by
a large number of its counterparts, the unit becomes lighter and less
sharply defined, ultimately going to zero mass while spreading out
completely at the enhan\c con locus.

Ref.~\cite{jpp} also went on to display a variety of familiar dual
situations in string theory (with related gauge theory physics) in
which the enhan\c con phenomena described above play a crucial
role. For this reason, and also because it is a genuinely new
mechanism by which string theory avoids an important class of
spacetime singularity\cite{singclass}, the enhan\c con deserves to be
better understood and characterised.

We show in this paper that enhan\c cons also arise naturally in
similar situations pertaining to large $N$ $SO(2N+1)$, $USp(2N)$ and
$SO(2N)$ $(p+1)$--dimensional gauge theories, and construct these new
classes using orientifolds. It is therefore clear that enhan\c cons
may be broadly classified into types $A$, $B$, $C$, and~$D$.  (There
is no natural $E$--type which has a smooth geometrical interpretation,
since the rank of those groups cannot be made arbitrarily large in
order to make contact with a supergravity discussion.) The latter
three types differ globally from type $A$ by having an extra $\ZZ$
identification, making them into $\IR{\rm P}^{4-p}\equiv S^{4-p}/\ZZ$
instead of $S^{4-p}$. All types can be distinguished locally by
examining their subleading behaviour in $N$.

As a concrete example, we shall focus in particular on $2{+}1$
dimensional gauge theory for $N$ large.  The reason that we focus on
this case is that we can explicitly write the relevant part of the
supergravity solution, using the fact that one of the relevant
orientifold 6--planes has a smooth M--theory realization as the
Atiyah--Hitchin manifold\cite{atiyah,sen}, while D6--branes are
related to Taub--NUT\cite{townsend}. While much of the structure of
the final result can be deduced on general grounds (the overall global
$\ZZ$ is the main feature), we show that a whole family of $1/N$
corrections can be completely characterised using the construction
that we present.

An overview is as follows: In the next section, we orient the reader
and set up our notation by reviewing the salient features of
ref.\cite{jpp}. In section~3, we discuss generally how orientifolds
yield the other types of enhan\c con. Crucially, we use a familiar
gauge theory fact to help us make a general statement about the result
of wrapping branes and orientifolds on $K3$.

In section 4, in preparation for the device of using the
Atiyah--Hitchin manifold to construct the eleven dimensional solution
for the $2+1$ dimensional $B,C,D$ gauge theory cases, we lift the
$A$--type case to M--theory and reconsider it in M--theory terms. In
particular, we observe that while in ten dimensions there is a
discussion of the geometry in terms of its constituents being
dynamical objects (D--branes), there is no analogous discussion
involving dynamical objects in M--theory. The geometry can be
discussed only in terms of the non--dynamical $K3$ and
multi--Taub--NUT.  Taking a probe to be an M2--brane fails to show the
enhan\c con, since the geometry cannot be constructed out of
them. Fortunately, long before one reaches the ``repulson''
singularity in the geometry, there is a sensible dual heterotic string
description (one of the duals in ref.\cite{jpp} (see also
refs.\cite{morten,senwitt})), where the geometry again has natural
brane probes revealing the enhan\c con. So we see again that the
enhan\c con mechanism resolves the physics of a supergravity
situation, this time by driving eleven dimensional supergravity back
to string theory.

In section 5, we show how to modify the discussion to make it
pertinent to the orientifolded enhan\c con, using the Atiyah--Hitchin
manifold combined with multi--Taub--NUT. On returning to string
theory, we study the supergravity solution, and extract the expression
for the enhan\c con radius and the leading $1/N$ correction which
follows from the orientifold's presence. We point out that an entire
class of $1/N$ corrections can be concisely summarised in terms of the
exponentially small differences between the smooth Atiyah--Hitchin
manifold and the (negative mass) Taub--NUT solution. Unfortunately, we
do not have such control over all of the corrections present in the
geometry.

We also present the result of a probe computation analogous to that
performed in ref.\cite{jpp} which yields the one--loop result for the
metric on a subspace of the Coulomb branch of moduli space for the
$B,C,D$ gauge theories. Again, they differ from the $SU(N)$ case by a
global $\ZZ$ action, and the  leading behaviour for the
$1/N$ corrections we computed. In all cases, there
is a related monopole moduli space problem, in the spirit of
refs.\cite{SWtwo,CH,hanany}.

\sect{The $A$--Type Enhan\c con} Consider wrapping $N$ coincident
D6--branes on a $K3$ surface of volume $V$.  This results in an
effective 2+1 dimensional object, with $-N$ units of D2--brane charge,
due to the interaction\cite{BSV}
\begin{equation}
-{\mu_6\over 48}\int C_{(3)}\wedge p_1({\cal R})
\label{induced}
\end{equation}
on the D6--brane world--volume. The precise value $-N$ comes about since
$\mu_6 = (2\pi)^{-6}
\alpha'^{-7/2}$, $\mu_2 = (2\pi)^{-2} \alpha'^{-3/2}$, 
${\cal R}=4\pi^2\alpha^\prime R$, and because for $K3$
\begin{equation}
p_1(R)\equiv{1\over 8\pi^2}R\wedge R\ ,
\end{equation}
integrates to $48$. We will call this wrong--sign D2--brane a
``D2*--brane''.  It preserves the same supersymmetries as a correct
sign D2--brane with the same orientation, and therefore is {\sl not}
an anti--D2--brane. It is useful to think of it as a brane which is
bound inside the D6--brane worldvolume, resulting from the curvature
of the $K3$. It is quite analogous to the (correct sign) D2--brane
which would be bound inside the worldvolume of a D6--brane if there
was a field theory instanton configuration, due to the term\cite{douglas} 
\begin{equation}
{\mu_6\over2}\int C_{(3)}\wedge{\cal F}\wedge {\cal F}\ ,
\end{equation} 
where ${\cal F}=2\pi\alpha^\prime F$. An instanton in the 6+1
dimensional gauge theory has $(8\pi^2)^{-1}\int F\wedge F=1$, and
consequently has the charge of a single D2--brane.  In the limit where
the instanton shrinks to zero size, there is a good
description\cite{edsmall} in the full string theory corresponding to a
fully localised pointlike D2--brane.

Similarly, one would recover pointlike D2*--branes from wrapping the
D6--branes if $K3$'s curvature was located at a finite number of
points, such as in an orbifold limit. This situation (no doubt) has
a good string theory description, and is worth investigating. In the
present case, the curvature of the $K3$ is distributed everywhere, and
correspondingly the D2*--branes are delocalised everywhere on it.

Imagine that the $K3$ surface lies in the $x^6,x^7,x^8,x^9$
directions, and that the remaining (unwrapped) part of the D6--brane
lies in the $x^0,x^1,x^2$ directions.  There is an $SU(N)$ gauge
theory on the $2+1$ dimensional worldvolume, with eight
supercharges. The gauge supermultiplet consists of a gauge field
$A_\mu$ and three scalars $\phi_i$, where $i=3,4,5$. The scalars
parameterise the positions of the D6--D2* system in the transverse
directions, $x^3,x^4,x^5$. This vector supermultiplet transforms in
the adjoint representation of $SU(N)$. The gauge theory has a scalar
potential of the form $\Tr[\phi_i,\phi_j]^2$.  Supersymmetric
solutions of the theory, giving a moduli space of vacua, may be found
by choosing vacuum expectation values (``vevs'') of the scalars such
that they are in the Cartan subalgebra of $SU(N)$. This breaks
$SU(N)\to U(1)^{N-1}$, giving the ``Coulomb branch'' of moduli space.

Classically, the moduli space is 
\begin{equation}
{\cal M}^N_{\rm cl}= \left.{\left(\IR^3\times S^1\right)\over
S_{N-1}}\right.^{N-1}\ ,
\end{equation}
where the $S^1$ factors represent the periodic scalars resulting from
dualising the gauge fields (recall that we are in $2+1$
dimensions). The $S_{N-1}$ is the Weyl group of $SU(N)$, which acts as
permutations of the $N-1$ eigenvalues of the $\phi$'s, which are now
in the Cartan subalgebra. $U(1)^{N-1}$ is the gauge symmetry on $N$
separated, but wrapped D--branes, where the extra $U(1)$ we would
naively expect corresponds to the overall centre of mass of the
system.

We will focus on the situation where all of the branes are coincident,
which is to say that the vev's of all of the fields are given the same
value, except for a complete set of four making a multiplet giving the
location of a probe brane in the background of all the others. In the
gauge theory, this is equivalent to focusing on a particular subspace
of the relative moduli space. In another, equivalent
description\cite{SWtwo,CH,hanany}, it is the four dimensional
$(1,N-1)$ subspace representing relative moduli space of the full
moduli space of $N$ $SU(2)$ monopoles; $N-1$ of them are coincident,
and one is separated. The classical moduli space is then
\begin{equation}
{\cal M}^{(1,N-1)}_{\rm cl}={\IR^3\times S^1}\ .
\label{monopole}
\end{equation}
One of the results of ref.\cite{jpp} (see below) is the computation of
the one--loop result for the metric on this moduli space. Here, we
shall compute a closely related version, representing a similar
subspace of the Coulomb branch of the $SO(2N+1)$, $USp(2N)$, or
$SO(2N)$ gauge theory. These will also have an interpretation as
multimonopole moduli spaces, where the monopoles are of an $SU(2)$
gauge theory with a $\ZZ$ identification.

For $gN$ large, ($g$ is the closed string coupling) we have a chance
of obtaining a good description of the geometry of the system in terms
of a ten dimensional type~IIA supergravity solution, which is
\begin{eqnarray}
ds^2 \* Z_2^{-{1\over2}} Z_6^{-{1\over2}} \eta_{\mu\nu} dx^\mu dx^\nu +
Z_2^{1\over2} Z_6^{1\over2} dx^i dx^i + V^{1\over2} Z_2^{1\over2} 
Z_6^{-{1\over2}} ds^2_{\rm K3} \
,\nonumber
\\
e^{2\Phi } \* g^2 {Z_2}^{1\over2}{ Z_6}^{-{3\over2}}\ , \nonumber\\
C_{\it (3)} \* g^{-1}(Z_2^{-1}-1) dx^0 \wedge dx^1 \wedge dx^2\ , \nonumber\\
C_{\it (7)} \* g^{-1}(Z_6^{-1}-1) dx^0 \wedge dx^1 \wedge dx^2 \wedge
dx^6 \wedge dx^7 \wedge dx^8 \wedge dx^9
\ . \label{sixtwo}
\end{eqnarray}
Here, $\mu,\nu=0,1,2$; $i=3,4,5$ and the $x^6,x^7,x^8,x^9$ directions
contain $ds^2_{\rm K3}$, the (unknown) metric of a unit volume $K3$.
The 345--harmonic functions representing the D2*-- and D6--branes
respectively are:
\begin{equation}
Z_2 = 1+\frac{r_2}{r}\qquad{\rm and}\qquad Z_6 =1+\frac{r_6}{r}\ ,
\end{equation}
(recall that the D2*'s are delocalised in $K3$), with
\begin{equation}
\quad r=|{\bar r}|, \quad{\bar r}\in \IR^3_{3,4,5}\ ,\quad r_2
=-\frac{ (2\pi)^4 g N \alpha'^{5/2} }{ 2V }\quad{\rm and}\quad r_6 =
\frac{gN\alpha'^{1/2}}{2} \ .
\end{equation}
The latter are written so as to give the masses of the BPS object
which is formed when we wrap a D6--brane to make the D6--D2* object:
\begin{equation}
\tau = \frac{N}{g} (\mu_6 V - \mu_2 )=\frac{N}{g}\mu_6(V-V_*)={N\over
g}\mu_2\left({V\over V_*}-1\right) \ ,
\label{tension2}
\end{equation}
where $V_*=(2\pi\sqrt{\alpha^\prime})^4$.

There are a number of things to note about this supergravity
solution. First, note that $g$ appears as the asymptotic value of the
string coupling far away from the core of the solution
($r\to\infty$). The actual string coupling in the interior of the
solution is given by the value of $e^\Phi$, as usual, and varies with
$r$. Similarly, the volume of $K3$ is a function of $r$: $V(r)=V
Z_2(r)/Z_6(r)$, which approaches $V$ asymptotically, and decreases,
becoming zero at the singularity $r=|r_2|$.

One of the key points noticed in ref.\cite{jpp} is that while a naive
examination of the supergravity solution shows an unsettling naked
singularity (the ``repulson''\cite{repulsive}) at $r=|r_2|$, this part
of the geometry is actually non--physical. The geometry should only be
taken at face value down to radius
\begin{equation}
r_{\rm e}={2V\over V-V_*}|r_2|\ .
\label{enhanced}
\end{equation}
This is the radius at which a
number of special things happen:
\begin{itemize}
\item The volume of $K3$ is equal to the special value
$V_*=(2\pi\sqrt{\alpha^\prime})^4$.
\item The 5+1 dimensional $K3$--compactified string theory has an R--R
sector $U(1)$ which becomes enhanced to an $SU(2)$ gauge symmetry.
\item A D6--D2* probe is a monopole of this $U(1)$, and becomes
massless at the enhanced symmetry point.  It also ceases to be
pointlike, and dissolves into the ``enhan\c con'' locus at~$r_{\rm
e}$, which is an $S^2$.
\end{itemize}
The interpretation of these and other facts uncovered in
ref.\cite{jpp} is that there are no brane sources for $r<r_{\rm e}$,
and therefore the supergravity solution inside that radius is simply
the trivial flat solution with no R--R fields switched on. The smooth
interpolating region between the two solutions in the neighbourhood of
the enhan\c con radius is described by the relatively innocuous (but
nonetheless interacting) monopole physics.

On the one hand, this situation represents another remarkable method
by which string theory rids itself of potentially troublesome
singularities, while on the other hand, it potentially teaches us
something about gauge theories. For example, in seeking for a limit in
which the gauge theory decouples from the rest of bulk physics, we
take $\alpha^\prime\to 0$ while holding fixed the 2+1 dimensional
gauge coupling given by:
\begin{equation}
g_{\rm YM}^2 = (2\pi)^{4} g \alpha'^{3/2}V^{-4}
\end{equation}
and hold fixed $U=r/\alpha'$. 
In this limit, it was found that the metric on the moduli space, as
seen by the D6--D2* probe, can be read off from the effective
Lagrangian for the monopole probe moving in the transverse space with
coordinates $(U,\theta,\phi,\sigma)$:
\begin{equation}
{\cal L}= f(U) \left({\dot U}^2
 +U^2{\dot \Omega}^2_2\right) +f(U)^{-1}
 \left({\dot\sigma} -{(N-1)\over{8\pi^2}}A_\phi{\dot\phi}\right)^2\
 ,\label{moduli}
\end{equation}
where\footnote{Note that we have inserted $N-1$ instead of the $N$
which appears in the supergravity solution~(\ref{sixtwo}) and also in
the probe result exhibited in ref.\cite{jpp}. Strictly speaking, there
are $N-1$ D6--D2* units being probed by one separated unit, giving $N$
in total. The difference is a $1/N$ effect, not considered in
ref.\cite{jpp}, but should be included here since we will later be
discussing a family of corrections at that order.}
\begin{equation}
f(U)={1\over 8\pi^2 g^2_{\rm YM}} \left(1-{g^2_{\rm YM}(N-1) \over
U}\right)\ ,\quad {\dot \Omega}^2_2={\dot
\theta}^2+\sin^2\theta{\dot\phi}^2\ ,
\end{equation}
with $0\leq\theta<0,\, 0\leq\phi<2\pi$ and $ U_{\rm
e}<U<\infty$. Here, $U_{\rm e} = g^2_{\rm YM}(N-1)$ and
$A_\phi=\pm1-\cos\theta$ is a $U(1)$ monopole potential.  The metric
in (\ref{moduli}) is the Euclidean Taub--NUT metric, with a negative
mass.  It is a hyperK\"ahler manifold, because $\nabla
f=\nabla{\times}A$, where $A=((N-1)/8\pi^2)A_\phi d\phi$. The
coordinate $\sigma$ is periodic with period $4\pi$, and is the dual of
the $U(1)$ centre--of--mass gauge field on the 2+1 dimensional
worldvolume of the monopole probe.

This result is completely in accord with the expectation from gauge
theory, being the one--loop result for the metric on moduli space, in
the special case where the $N-1$ coordinates parameterising the Cartan
subalgebra are chosen to be equal, corresponding to making all of the
branes coincident. The enhan\c con is at $U=U_{\rm e}$, and
corresponds to the Landau pole, representing in gauge theory the place
where the one--loop correction makes the gauge coupling diverge.

In the equivalent monopole language, this is (an approximation to) the
metric on a subspace ${\cal M}^{(1,N-1)}$ (described above
eqn.~(\ref{monopole})) of the full moduli space of $N$ $SU(2)$
monopoles. This $SU(2)$ is the enhanced gauge symmetry from whence
comes the name ``enhan\c con''.

There are exponential corrections to this metric which will remove the
singular behaviour and complete it into a smooth hyperK\"ahler
manifold, ${\cal M}^{(1,N-1)}$. This space generalises the
Atiyah--Hitchin manifold, which is the metric on the relative
two--monopole moduli space ${\cal M}^{(1,1)}$ which governs the case
of $SU(2)$ gauge theory\cite{SWtwo}.

A natural question arises about the nature of the story for the case
where one studies gauge groups other than $SU(N)$. It is
straightforward to construct gauge groups $SO(2N)$, $SO(2N+1)$, and
$USp(2N)$ in perturbative string theory by combining D--branes with
orientifolds. Studying the wrapping of such a system on $K3$ should
therefore be our first step in answering the question. Let us do that.

\section{Including Orientifolds}

On general grounds, one expects a similar story to that which was
constructed above, as all of the constituent features which are
present to make the physics work as it should are still present after
we insert an orientifold six--plane (O6--plane) parallel with the
D6--branes. Of course, the details of precisely where the enhan\c con
is located (corresponding to where in $\IR^3_{345}$ the $K3$ volume
reaches the value $V_*$) will be modified, but only at subleading
order in $N$.  Globally, the orientifold will also place a $\ZZ$
identification on $\IR^3_{345}$ ($\ZZ$ acts by multiplying each of
$x^3,x^4,x^5$ by $-1$), turning the $S^2$ of the enhan\c con into
$\IR{\rm P}^2\equiv S^2/\ZZ$. The basic problem is to understand the
nature of the supergravity solution in the presence of the
orientifolds, which we will do below in a particular case.

First, let us understand the physics of the perturbative string theory
description, containing the weakly coupled gauge theory. For small
$gN$, we have $N$ D6--branes, and an O6--plane parallel to them.  This
gives a gauge group $SO(2N+1)$, $USp(2N)$ or $SO(2N)$. In the latter
case, the O6--plane has negative charge, equal to $-2\mu_6$ and is
often denoted O6$^-$.  We can obtain the former case by trapping a
half D6--brane on the O6$^-$--plane: this combination is often
referred to as an $\widetilde{\rm O6}$--plane, with charge
$-3\mu_6/2$. In the middle case, the O6--plane has charge $+2\mu_6$
and is written O6$^+$.  To be concise, we will use the symbol $\alpha$
to denote these O6--charges, measured in D2--brane units.  It takes
the values $\alpha=-3/2,+2,-2$, respectively.

We now wrap the whole system on $K3$. This results in the induced
D2*--branes as described above, but with an additional
contribution. This is due to a curvature coupling, this time on the
world--volume of the O6--plane\cite{ocurve}, similar to
eqn.(\ref{induced}).  The couplings are different in each case
${\widetilde {\rm O6}}$, O6$^+$, O6$^-$:
\begin{equation}
-{\mu_6\over 32}\int C_3\wedge p_1({\cal R})\ ,\quad
-{5\mu_6\over 48}\int C_3\wedge p_1({\cal R})\ ,\quad
-{\mu_6\over 48}\int C_3\wedge p_1({\cal R})\ ,
\label{inducedtwo}
\end{equation}
which, after wrapping on $K3$ will induce some $C_{(3)}$ charge,
$\beta$, which in D2--brane units, is respectively
$\beta=-3/2,-5,-1$. This will modify the contribution to the effective
amount of D2*--brane present\footnote{The temptation to interpret
these extra charges as induced wrong sign O2--planes should, we
believe, be firmly resisted. First of all, the resulting charges are
hard to interpret, given the existing types of O2--brane already
known. Secondly, one would have to insert a $\ZZ$ identification on
the $K3$ part of the spacetime, which is hard to justify as the result
of a smooth wrapping process. The most economical interpretation is
the one presented here.}.

Note that we can introduce extra (correct sign) D2--branes parallel to
the D2*--branes into the story while preserving the eight
supercharges. Open strings stretching between these new branes and the
wrapped system play the role of extra hypermultiplets in the 2+1
dimensional gauge theory. For $M$ D2--branes, we have $M$ species of
such hypermultiplets. For consistency, in the presence of the
orientifold, these D2--branes will have the {\sl opposite} orientifold
projection acting on them from that acting on the D6's, as follows
from T--duality to the situations studied in
refs.\cite{edsmall,joeeric}.

\begin{figure}[ht]
\hbox{\epsfxsize=2.7in\epsfbox{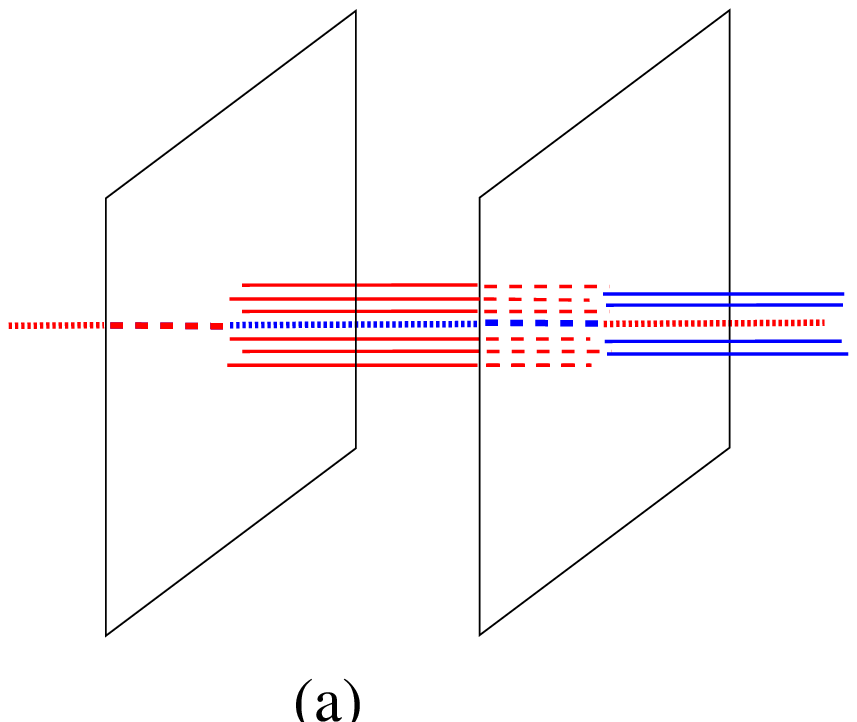}\hskip0.4in
{\epsfxsize=2.4in\epsfbox{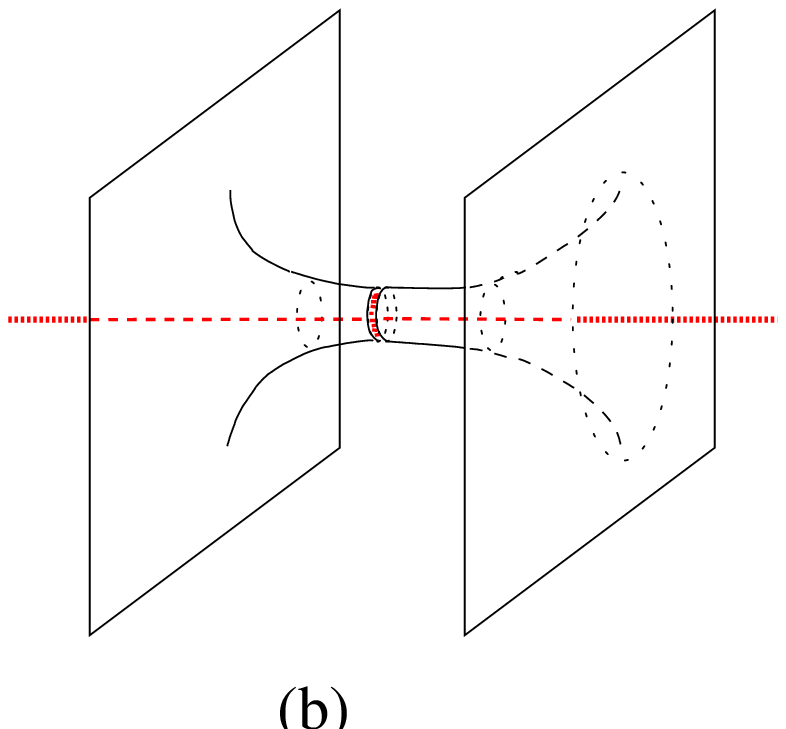}}}
\caption{\small D$(p+1)$--branes ending on NS5--branes, in the
presence of an O$(p+1)^-$--plane (central dotted line). (a) The $N$
branes and their images in the interior give an $SO(2N)$ gauge
group. On the exterior, $M$ branes and their images supply matter
hypermultiplets, and carry a gauge group $USp(2M)$, as the orientifold
plane changes sign when it goes through an NS5--brane. (b) At large
$gN$, the fivebranes will be bent, touching at an $\IR{\rm
P}^{4-p}{=}S^{4-p}/\ZZ$, (whose double cover is a circle in the
figure), the $D$--type enhan\c con, where the NS5--branes carry an
enhanced gauge symmetry. (In case (b), for clarity, no matter branes
are shown.)}
\label{bendy}
\end{figure}

So, for example, while there is an $SO(2N)$ (or $USp(2N)$) gauge
symmetry on the D6--branes, the $M$--flavour sector has a $USp(2M)$
(or $SO(2M)$) symmetry. This is indeed correct from the perspective of
gauge theory\cite{gaugefact}, and this fact has featured in the
physics of orientifolds before. In ref.\cite{ejs}, it was shown to
correspond to the phenomenon that the orientifold must change its sign
when it passes through an NS5--brane. Actually, one of the dual
realizations of the enhan\c con story involves
NS5--branes\cite{jpp}. We display it with the inclusion of the
orientifold in figure~\ref{bendy}, where our case here is $p=2$.  The
D3--branes in the interior are dual to the D6--D2* units, while the
those on the exterior (supplying matter multiplets) are dual to the
ordinary D2--branes.  The orientifold runs through the whole system,
having a minus sign on the interior (giving $SO(2N)$) and a plus sign
on the exterior, giving $USp(2M)$.

So we see that the sign flip of the O3--plane on either side of the
NS5--brane is dual to the fact that the O6--plane has the opposite
projection on the D2's from that on the D6's.  Of course, this
discussion clearly generalises to all D$p$--branes and D$p$*--branes
with orientifolds, and the dual situation involving D$(p+1)$--branes
stretched between NS5--branes with an O$(p+1)$--plane passing through.

In this way, we see that we can consistently construct wrapped
D--brane and orientifold systems giving gauge groups $SO(2N+1)$,
$USp(2N)$, and $SO(2N)$.  The amount of D2*--branes induced from the
wrapping is modified from $-N$ (for the $SU(N)$ case) to $-N{-}3/2$,
$-N{-}5$ and $-N{-}1$, respectively.  The changes are to be thought of
as $1/N$ corrections to the original case, and are different for each
type of orientifold. We can now consider taking $gN$ large, and expect
that the phenomena which occurred for $SU(N)$ will happen again,
giving an enhan\c con for each case. We shall name the types of
enhan\c con which can occur in each situation the $A$--type (for
$SU(N)$), $B$--type (for $SO(2N+1)$), $C$--type (for $USp(2N)$) and
$D$--type ($SO(2N)$).

Of course, there is no natural definition of an $E$--type, for (at
least) two reasons: There is no known perturbative way to make
$E_{6,7,8}$ gauge symmetry with D--branes, and furthermore, the
enhan\c con as a smooth geometric object is a large $N$ phenomenon,
which is incompatible with the fact that the maximum rank of the
exceptional groups is eight.

The next step in our story will be to write down the geometry
corresponding to the large $gN$ physics of the wrapped system of
D--branes and orientifolds.  The observation that we shall use to
achieve this is the fact that for the O6$^-$ case (giving 2+1
dimensional $SO(2N)$), the supergravity geometry of the system can be
written down accurately enough for us to make progress. Along the way,
we will see that we can study cases $SO(2N+1)$ and $USp(2N)$
accurately enough for our purposes using similar techniques.

\section{Uplifting the Enhan\c con}

Before we proceed to the new types, let us pause for a moment to
consider the $A$--type enhan\c con story in eleven dimensional terms.
Recall that the metric of the Taub--NUT space, made into an eleven
dimensional supergravity solution (by adding $\IR^{6,1}$ for the
world--volume directions) is (defining an eleventh direction
$\psi=x^\natural/16m$):
\begin{eqnarray}
&&ds^2_{11}=-dt^2+dx_1^2+dx_2^2+dx_6^2+dx_7^2+dx_8^2+ dx_9^2\\
&&\hskip2cm+F(r)(dr^2+r^2d\Omega_2^2)+F^{-1}(r)\left(d\psi+C_\phi
d\phi\right)^2\ ,\nonumber
\end{eqnarray}
where, with  $0\leq\psi<4\pi$,
\begin{equation}
d\Omega_2^2=d\theta^2+\sin^2\theta d\phi^2\ , \quad F=1+{4mN\over r}\
,\quad C_\phi=4mN\cos\theta\ .
\end{equation}

Reducing along the $\psi$--circle, the relation between eleven
dimensional metrics and ten dimensional type IIA fields
is\cite{wittenvarious}:
\begin{equation}
ds_{11}^2=e^{-{2\over3}\phi}ds^2_{10}+e^{{4\over3}\phi}(d\psi+C_{(1)})^2\
,
\label{uplift}
\end{equation}
and so we recover the now standard fact\cite{townsend} that Taub--NUT
 corresponds to a familiar ten--dimensional solution:
\begin{eqnarray}
&&ds^2_{10}=Z_6^{-{1\over2}}(-dt^2+dx_1^2+dx_2^2+dx_6^2+dx_7^2+dx_8^2+
dx_9^2)\nonumber\\
&&\hskip9cm+Z_6^{1\over2}(dr^2+r^2d\Omega_2^2)\nonumber\\ &&Z_6=F(r)\
;\qquad e^{\phi}=Z_6^{-{3\over4}}(r)\ ,\qquad C_\phi=4mN\cos\theta\ ,
\end{eqnarray}
which is precisely the D6--brane solution, if we identify $4mN=r_6$
(and set the asymptotic value of the dilaton to $\log g$.)  The
one--form potential $C_{(1)}=C_\phi d\phi$ can be Hodge--dualised in
ten dimensions to give an electric source for $C_{(7)}$ of precisely
the form given in eqn.~(\ref{sixtwo}).

Turning to the enhan\c con, by using the prescription of
eqn.~(\ref{uplift}), supplemented with a direct uplift of the
three--form potential $C_{(3)}$ to give the components of the eleven
dimensional three--form $A_{(3)}$, it is easy to write an eleven
dimensional solution for the uplifted D6--D2* system:
\begin{eqnarray}
&&ds^2_{11}={\widetilde
Z}_2^{-{2\over3}}(-dt^2+dx_1^2+dx_2^2)+{\widetilde Z}_2^{{1\over3}}
{\widetilde V}^{1\over2}ds_{\rm K3}^2\nonumber\\ &&\hskip2cm
+{\widetilde Z}_2^{{1\over3}}\left[{\widetilde
Z}_6(dr^2+r^2d\Omega_2^2) +{\widetilde Z}_6^{-1}\left(d\psi+C_\phi
d\phi\right)^2\right]\ ,\nonumber\\ {\rm with}&&\quad
A_{(3)}=\left({\widetilde Z}_2^{-1}-1\right)dx^0\wedge dx^1\wedge dx^2
\ ,
\label{msoln}
\end{eqnarray}
\begin{equation}
{\rm where}\quad {\widetilde V}=g^2V\ ,\quad {\widetilde Z}_2=gZ_2\
,\quad {\widetilde Z}_6=g^{-1}Z_6\ .
\end{equation}

It is interesting to contrast the interpretation of this solution with
the ten dimensional discussion. Recall that from the point of view of
ten dimensions, there is the geometry of $K3$, accompanied by
D6--branes wrapped on it. The wrapping induced some D2*--branes,
completely delocalised in the $K3$. We were able to probe the geometry
of the supergravity solution~(\ref{msoln}) with one of its basic
constituents, a single D6--D2* BPS object.

{}From the point of view of the eleven dimensional supergravity
solution, everything is geometry: there are no branes here.  The
Taub--NUT part lies in the 345$\natural$ directions, while $K3$ lies
in the 6789 directions.  Together, they act as a source for the
three--form potential $A_{(3)}$, due to the supergravity
term\cite{elevencouples}:
\begin{equation}
\int A_{(3)}\wedge X_8\ ,\qquad {\rm with}\quad X_8={1\over24}
\left(p_2-{1\over4}p_1\wedge p_1\right)\ .
\label{minduced}
\end{equation}
Given that for Taub--NUT of charge $N$, $p_1=2N$ and, as stated
before, for $K3$ we know that $p_1=48$, we get $-N$ units of $A_{(3)}$
charge, as the solution shows. This fits, as (\ref{minduced}) is the
M--theory ancestor of the brane world--volume term~(\ref{induced}).

Sadly, there is no natural extended dynamical object which we can use
as a candidate for the basic constituent of the geometry. Thus, we
cannot perform a world--volume probe computation to deduce the true
geometry.  It is tempting to read the ${\widetilde Z}_2$ part of the
geometry as representing a ``wrong sign'' M2--brane which is otherwise
dynamical, (perhaps restoring a $1/r^3$ behaviour to make it also
localised in the $x^\natural$ direction.)  Unfortunately, this cannot
work. The putative M2*--brane necessarily would have negative tension
at all locations in $\IR^3_{345}$, and since there is no larger
wrapped brane with positive tension to combine it with to made a
positive tension object, we cannot write a sensible worldvolume
action. Of course, a probe computation with a correct sign M2--brane
(which preserves the same amount of supersymmetry) gives a sensible
result: simply the pure (with mass parameter of $+4mN=r_6$) Taub--NUT
metric, as it should, with no sign of either enhan\c con or repulson.
This is in accord with our expectation that the repulson (still
present at $r=|r_2|$) is an artifact, while the enhan\c con should be
invisible to an M2--brane because its world--volume theory does {\sl
not} relate to the $SU(N)$ gauge theory.

To get at the enhan\c con, there is a pertinent supergravity question
to be asked all the same: Can we envision a supergravity mechanism by
which the troublesome repulson singularity at $r=|r_2|$ is avoided? In
the string theory situation, we saw that the $K3$ reached the natural
value $V_*=(2\pi\sqrt{\alpha^\prime})^4$, before it reached its
singular value (zero), and the physics of the enhanced gauge symmetry
took over.  Is there a special value for the volume of $K3$ in this
case?  Here, the natural length scale is of course set by the Planck
length, $\ell_{11}=g^{1/3}\ell_s$, which the system again reaches
before the singular value of zero. Once $K3$ has shrunk to that size,
we should search for a better description than eleven dimensional
supergravity. The alternative to using the full (unknown) M--theory is
to search for a dual description. Happily, eleven dimensional
supergravity on such a small $K3$ is well
described\cite{wittenvarious} by the heterotic string on $T^3$. The
distinguished $\psi$--circle which is fibred to make the Taub--NUT
joins the rest to become a~$T^4$, and the Taub--NUT structure becomes
a ``warped''\cite{jkkm} (not ``wrapped'') NS--fivebrane/Kaluza--Klein
monopole structure giving rise to a monopole membrane whose mass goes
to zero at an $SU(2)$ enhanced point of the torus\cite{jpp,morten}
(see also ref.\cite{senwitt}).  So we see that again, stringy physics
(now heterotic) takes over before we get to the repulson radius, and
repairs the geometry with the same $SU(2)$ physics that we saw in type
IIA string theory.

\section{The Orientifolded Enhan\c con}
Just as the D6--brane has an interpretation as the Taub--NUT spacetime
upon going to low energy M--theory (eleven dimensional supergravity),
in a similar fashion, the O6$^-$--plane becomes\cite{sen} the
Atiyah--Hitchin manifold\cite{atiyah}, described by a metric:
\begin{eqnarray}
&&ds^2_{11}=-dt^2+dx_1^2+dx_2^2+dx_6^2+dx_7^2+dx_8^2+ dx_9^2
\label{atiyah} \\
&&\hskip3cm+f(\rho)dr^2+8m^2\left(a^2(\rho)\sigma_1^2
+b^2(\rho)\sigma_2^2+c^2(\rho)\sigma_3^2\right)\ ,\nonumber
\end{eqnarray}
where
\begin{eqnarray}
&&\sigma_1=-\sin\psi d\theta+\cos\psi\sin\theta d\phi\ ,\qquad
\sigma_2=\cos\psi d\theta+\sin\psi\sin\theta d\phi\nonumber \\
&&\sigma_3= d\psi+\cos\theta d\phi\ ,\qquad \rho={r\over 8m}\ ,
\qquad\psi={x^\natural\over 16m}\ ,
\end{eqnarray}
and the functions $f,a,b,c,d$ are given in terms of elliptic integrals
in ref.\cite{gibbonsmanton}.  There is also an identification by:
\begin{equation}
(r,\theta,\phi,\psi)\to(r,\pi-\theta,\pi+\phi,-\psi)\ ,
\label{identification}
\end{equation}
which, in terms of the coordinates $(x^3,x^4,x^5)$ and the
M--direction $x^\natural$, is simply a multiplication by a minus sign
on all directions. The displayed metric (\ref{atiyah}) has a conical
singularity at $r=8\pi m$.  The space made by imposing the $\ZZ$
identification (\ref{identification}) is the Atiyah--Hitchin space,
and it is free of conical singularities.

While a closed form for the metric cannot be written, for large $r$
the metric becomes\cite{gibbonsmanton}
\begin{eqnarray}
&&ds^2_{11}=-dt^2+dx_1^2+dx_2^2+dx_6^2+dx_7^2+dx_8^2+ dx_9^2+\nonumber\\
&&\hskip1.5cm+G(r)(dr^2+r^2d\Omega_2^2)+G^{-1}(r)\left(d\psi+C_\phi
d\phi\right)^2\ ,
\end{eqnarray}
where
\begin{equation}
G=1-{16m\over r}\ ,\qquad C_\phi=16m\cos\theta\ ,
\end{equation}
which we recognise as the metric for Taub--NUT, but with a negative
mass. Clearly, it can be reduced to ten dimensions in the same way as
before, and we see \cite{sen} that it has $-2$ units of D6--brane
charge, which is in accord with our knowledge of the charge of an
O6$^-$--plane from perturbative string theory. (The actual appearance
of $-16m$ in the metric instead of $-8m$ follows from the fact that
the displayed metric is the double cover of the actual solution:
recall that we must divide by the $\ZZ$ action.)

Now we are in a position to construct the geometry which gives rise
the the $D$--type enhan\c con.  We simply combine the geometry of the
Atiyah--Hitchin manifold with that of $N$ coincident--centred
Taub--NUT.  The exact smooth metric certainly exists (the $N=1$ case
is known, and is Dancer's manifold\cite{dancer}), but we need not be
able to write it exactly to get at the physics we require. The radius
at which the enhan\c con appears can be tuned to be arbitrarily large
by making $N$ as large as we like, so we can rest assured that if we
take the approximate expression for the Atiyah--Hitchin manifold, we
can capture the essential physics for large $N$.

Once we relax the condition of exactness, and focus on the large $r$
part of the solution, we can include the cases of the $B$-- and
$C$--type enhan\c cons. While a precise relation to a cousin of the
smooth Atiyah--Hitchin$+$Taub--NUT geometry is not known, at
large~$r$, the difference is immaterial, as only the leading behaviour
is needed to characterise the enhan\c con at large enough $N$.  We can
simply use the same supergravity solution as before, but with
different numbers inserted into the $1/N$ corrections to the harmonic
functions.

It is clear therefore, that for all cases our solution can be written
(for large enough~$r$) in the covering space in the precise form of
eqn.~(\ref{msoln}), but with the replacement of ${\widetilde Z}_2$ and
${\widetilde Z}_6$ by (respectively):
\begin{equation}
{\widetilde Z}^\prime_2=g\left(1-{2|r_2|(1-\beta/N)\over r}\right)
\quad {\rm and}\quad {\widetilde
Z}^\prime_6={1\over g}\left(1+{2r_6(1+\alpha/N)\over r}\right)\ .
\label{corrected}
\end{equation}
Here\footnote{It is amusing to note that the sum $\alpha+\beta$ is the
 same in each case. We do not know if this has any physical
significance.  Later, in eqn.~(\ref{newradius}), we shall see that it
is $\alpha-\beta$ which controls the leading $1/N$ correction to the
enhan\c con in each case. Were it the sum which appeared, we would
have had a remarkably universal result.}  $\beta=-3/2,-5,-1$, and
$\alpha=-3/2,+2,-2$ for types $B, C, D$, respectively.  We have
deduced ${\widetilde Z}^\prime_2$'s asymptotic form\footnote{While we
know (in the $D$--type case) precisely how the harmonic function of
${\widetilde Z}^\prime_6$ gets corrected into the smooth
Atiyah--Hitchin+Taub--NUT solution, we do not know how ${\widetilde
Z}^\prime_2$, which owes its presence to the $K3$ part of the eleven
dimensional geometry, gets corrected. In its current form, it must be
there in order to measure the correct mass and charge at large $r$,
but the small $r$ details are unknown to us.}  from the fact that it
must give the correct induced D2*--brane mass and charge at large $r$
in the string theory limit.

This should be taken to mean the metric on the covering space of our
solution, and we must divide by the $\ZZ$ action in order to
reconstruct the correct solution, as before. This also accounts for
the factors of two we have inserted into the harmonic
functions. Notice that the contribution to the harmonic functions of
(what will become) the orientifolds is simply a~$1/N$ correction to
the geometry. This will turn into part of the family of $1/N$
corrections to the location and shape of the enhan\c con locus, once
we return to string theory.

The final step is clear. We return to type IIA string theory by
reducing on the $\psi$--circle, recovering a supergravity solution
representing the large $gN$ geometry of system of wrapped D6--brane
and and O6$^-$--plane, as promised in section 2.  The solution is
simply the geometry (\ref{sixtwo}) with $Z_2$ and $Z_6$ replaced by
their $1/N$ corrected counterparts in (\ref{corrected}) with the
factors of $g$ and $1/g$ removed. Crucially, there is a $\ZZ$
identification on the $(x^3,x^4,x^5)$ directions, making it globally
distinct from the $A$--type case, in addition to the different
structure of the subleading behaviour in $N$.

Again, in string theory, the natural object to construct this geometry
out of is the D6--D2* at large $gN$, now in the presence of an
orientifold, and we may examine the nature of the geometry as seen by
the probe by a computation precisely along the lines of
ref.\cite{jpp}.  The structure of the computation is almost identical
to that carried out there, and we refer the reader to that work for
the details. A crucial difference is that we are working on the {\sl
covering space} of the actual geometry, and so we should insert a
mirror image of the probe at the image position obtained by reflecting
through the orientifold fixed point.  The result is structurally
identical:
\begin{equation}
{\cal L}=F^\prime(r)\left({\dot
r}^2+r^2{\dot\Omega}^2_2\right)+F^\prime(r)^{-1}\left({\dot s}/2
-\mu_2C_\phi{\dot \phi}/2\right)^2\ ,
\end{equation}
where now
\begin{equation}
F^\prime(r)={1\over2g}(\mu_6VZ^\prime_2-\mu_2Z^\prime_6)\ ,
\end{equation}
with
\begin{eqnarray}
Z^\prime_2&=&\left(1-{2|r_2|(1-(\beta+1)/N)\over r}\right) \quad{\rm
and}\quad \nonumber\\ Z^\prime_6&=&\left(1+{2r_6(1+(\alpha-1)/N)\over
r}\right)\ ,
\label{corrections}
\end{eqnarray}
where we have shifted $N$ to $N-1$ to represent separating off the
probe (see footnote 1).  Here, $s$ is the fourth modulus obtained by
dualising the world--volume centre of mass gauge field.  The location
$r^\prime_{\rm e}$ of the $D$--type enhan\c con can be read off as the
place in $r$ where the mass of the probe becomes zero (equivalent to
$V(r^\prime_{\rm e})=V_*$):
\begin{equation}
r^\prime_{\rm e}={2V\over V-V_*}|r_2|\left(1-{\gamma\over N}\right)\ ,
\label{newradius}
\end{equation}
with 
\begin{equation}
\gamma=-\left({\alpha-\beta-2\over2}\right)=1\ ,-{5\over 2}\ ,{3\over2} \,
\end{equation}
in each case $B, C, D$. (The analogous expression for the $A$ case
---with the $1/N$ correction from separating off the probe
({\it c.f.} eqn.~(\ref{enhanced}))--- has $\gamma=1$. Note that for
case $B$ the effect of the O6--plane is precisely cancelled by the
effect of the D2*--brane contribution which is produces from wrapping,
giving the same leading $1/N$ contribution as for type $A$.)

Correspondingly, when we take the limit where we decouple the gauge
theory with $\alpha^\prime\to0$ holding $g^2_{\rm YM}$ fixed, we
recover the prediction for the metric on the moduli space of the
 gauge theory at large $N$ (in the coincident limit):
\begin{equation}
{\cal L}= f^\prime(U) \left({\dot U}^2
 +U^2{\dot \Omega}^2_2\right) +f^\prime(U)^{-1}
 \left({\dot\sigma} -{N(1-\gamma/N)\over{8\pi^2}}A_\phi{\dot\phi}\right)^2\
 ,\label{moduliprime}
\end{equation}
where
\begin{equation}
f^\prime(U)={1\over 8\pi^2 g^2_{\rm YM}} \left(1-{g^2_{\rm YM}N \over
U}\left(1-{\gamma\over N}\right)\right)\ .
\label{neweff}
\end{equation}
This is the one--loop expression for the metric on moduli space for
the $SO(2N+1)$, $USp(2N)$ or $SO(2N)$ 2+1 dimensional gauge theory.
On general grounds, the classical moduli space has the geometry
\begin{equation}
{\cal M}_{\rm cl}=\left.{\left(\IR^3\times S^1\right)\over S_N\times
\ZZ}\right.^N
\end{equation}
where there is a natural $\ZZ$ action reflecting the $N$ eigenvalues
into (minus) themselves.  In the subspace where we set all the vev's
(but four) to be equal, we are reduced to
\begin{equation}
{\cal M}_{\rm cl}={\IR^3\times S^1\over {\ZZ}}
\end{equation}
for the classical moduli space. Our metric above, with the $\ZZ$
action (imposed, recall, for smoothness of the Atiyah--Hitchin
manifold representing the O6$^-$--plane), is the one--loop expression
for the metric on the full moduli space. 

Finally, we point out that once again, these results have a dual
interpretation as an approximate result for the metric on moduli space
of $N$ monopoles\cite{SWtwo,CH,hanany}.  This time, they are monopoles
of a spontaneously broken $SU(2)$ theory which has an identification
by $\ZZ$, which can be understood as follows\footnote{See
ref.\cite{zeetwo} for comments on such theories in a closely related
stringy context.}: The relation between the moduli space of $2+1$
dimensional gauge theories and that of monopoles is readily seen in
the string realization of such theories by D3--branes stretched
between NS5--branes\cite{hanany}. The spontaneously broken $SU(2)$
lives on the world--volume of the NS5--branes. The ends of the
D3--branes in the NS5--brane worldvolumes are the monopoles. We need
only look at the orientifolded version of that picture, drawn in
figure~1, to see the origin of the $\ZZ$ action on the $SU(2)$
theory. By passing through the world--volume of the NS5--branes, the
O3--plane places a spacetime $\ZZ$ identification on the $SU(2)$ gauge
theory.

\section{Closing Remarks}
The enhan\c con locus which appears in the study of spacetime geometry
associated to $SU(N)$ $(p+1)$--dimensional gauge theory (at large $N$)
with eight supercharges has three natural counterparts: Those
pertaining to $SO(2N+1)$, $USp(2N)$ and $SO(2N)$ gauge theory. The
four classes deserve to be called types $A$, $B$, $C$, and $D$. (There
is no natural $E$--type which has a smooth geometrical interpretation,
since the rank of those groups cannot be made arbitrarily large.)

We presented the general scenario for the case of $(p+1)$--dimensions
and exhibited and studied the orientifolded enhan\c con for the case
of 2+1 dimensional gauge theory. Guided by the case of the $D$--type,
where the fact that the O6$^-$--plane has a known eleven dimensional
supergravity description in terms of the Atiyah--Hitchin manifold, we
were able to study aspects of all three new types: While the
Atiyah--Hitchin manifold cannot be written explicitly, it reduces to
(negative mass) Taub--NUT at large $r$ (up to exponentially small
corrections in $r$) which was enough for us to study explicitly the
relevant features of the supergravity solution which results from
placing many D6--branes and an O6--plane on $K3$. This
multi--Taub--NUT solution can also be reliably modified to capture the
local asymptotic behaviour of the $B$-- and $C$--type cases. The
Atiyah--Hitchin structure imposes a global $\ZZ$ identification on the
entire geometry. Correspondingly, we found that there is a global
$\ZZ$ identification inherited by the enhan\c con locus, making the
enhan\c con a natural $\IR {\rm P}^2\equiv S^2/\ZZ$ geometry, in
contrast to the $S^2$ geometry of the $A$--type. (We should also note
that we observed that in all cases $A, B, C,$ or $D$, the apparent
repulson singularity in eleven dimensional supergravity is naturally
removed; not by M--branes, but by being forced back to ten dimensional
heterotic string theory because the $K3$ becomes small. The heterotic
string phenomena dual to the enhan\c con\cite{jpp} then take
over the description.)

We displayed some leading $1/N$ corrections to the location of the all
three types of orientifolded enhan\c con, as compared to the location
of the $A$--type, and hence also the $1/N$ corrections to the
one--loop metric on moduli space.  Note that the $A$--type enhan\c con
already has a series of exponential corrections of the form
$\exp(-1/g^2_{\rm YM})$.  On general grounds, our new types have a
similar class of corrections, which can be phrased in terms of field
theory instanton corrections\cite{SWtwo,khoze}, and equivalently, in
terms of corrections from  D1--brane
world--sheets\cite{hanany,ahn}.

It is amusing to note that the $1/N$ corrections we studied here,
which are of a different type, can all be written in terms of
exponential corrections too.  This follows from the fact that the part
of the geometry of the Atiyah--Hitchin (--like) manifold that we
neglected in writing the explicit supergravity solution is a series of
exponential corrections in $r$.  These corrections should also have an
interpretation in terms of an instanton problem. Perhaps one can
always organise the exponential corrections to these geometries in
terms of structures reminiscent of the geometry of the Atiyah--Hitchin
manifold, regardless of whether they are non--perturbative
in~$g^2_{\rm YM}$ or~$1/N$.

\section*{Acknowledgements}
We would like to thank Peter Bowcock, George Papadopoulos, Simon Ross,
Douglas Smith, Paul Sutcliffe and David Tong for comments and  discussions.



\begin{thebibliography}{99}
\baselineskip=17pt
\bibitem{jpp}C. V. Johnson, A. Peet and J. Polchinski, hep-th/9911161, 
to appear in Phys. Rev. {\bf D}.
\bibitem{repulsive} K.~Behrndt,
Nucl.\ Phys.\ {\bf B455}, 188 (1995)
hep-th/9506106;\hfill\\
R.~Kallosh and A.~Linde,
Phys.\ Rev.\ {\bf D52}, 7137 (1995)
hep-th/9507022;\hfill\\
M.~Cveti{\v{c}} and D.~Youm,  
Phys.\ Lett.\ {\bf B359}, 87 (1995)
hep-th/9507160.

\bibitem{singclass}As pointed out by R. C. Myers and also
A. Strominger, as cited in ref.\cite{jpp};
J. Polchinski and M. Strassler, in progress;\hfill\\
S.S.~Gubser, K.~Pilch and N.P.~Warner, in progress;
\hfill\\ For recent discussions, 
some with relation to gauge theory, see:\hfill\\
S.~S.~Gubser,
hep-th/0002160;\hfill\\
I.~R.~Klebanov and A.~A.~Tseytlin,
hep-th/0002159;\hfill\\K.~Behrndt and M.~Cvetic,
hep-th/0002057;\hfill\\
M.~Cvetic, H.~Lu and C.~N.~Pope,
hep-th/0002054;\hfill\\
T.~Takayanagi,
hep-th/9912157;\hfill\\
I.~Bakas, A.~Brandhuber and K.~Sfetsos,
hep-th/9912132.


\bibitem{atiyah}
M.~F.~Atiyah and N.~J.~Hitchin,
Phys.\ Lett.\  {\bf A107}, 21 (1985);\hfill\\
M.~F.~Atiyah and N.~J.~Hitchin,
Phil.\ Trans.\ Roy.\ Soc.\ Lond.\  {\bf A315}, 459 (1985).

\bibitem{sen}A.~Sen,
JHEP {\bf 9709}, 001 (1997), hep-th/9707123;\hfill\\ A.~Sen,
JHEP {\bf 9710}, 002 (1997), hep-th/9708002

\bibitem{townsend}P. Townsend, Phys.Lett. B350 (1995) 184, hep-th/9501068.


\bibitem{morten} M.~Krogh,
JHEP 9912 (1999) 018, hep-th/9911084.

\bibitem{senwitt}
A.~Sen,
Adv.\ Theor.\ Math.\ Phys.\  {\bf 1}, 115 (1998), hep-th/9707042;\hfill\\
E.~Witten,
hep-th/9909229.

\bibitem{SWtwo} N.~Seiberg and E.~Witten,
hep-th/9607163.
\bibitem{CH}G.~Chalmers and A.~Hanany,
Nucl.\ Phys.\ {\bf B489}, 223 (1997)
hep-th/9608105.
\bibitem{hanany}A.~Hanany and E.~Witten,
Nucl.\ Phys.\ {\bf B492}, 152 (1997) hep-th/9611230.


\bibitem{BSV} M.~Bershadsky, C.~Vafa and V.~Sadov,
Nucl.\ Phys.\ {\bf B463}, 398 (1996) hep-th/9510225;\hfill\\ M.~Green,
J.A.~Harvey and G.~Moore,
Class.\ Quant.\ Grav.\ {\bf 14}, 47 (1997)
hep-th/9605033.

\bibitem{douglas}M. R. Douglas, hep-th/9512077.
\bibitem{edsmall}E. Witten, Nucl.Phys. B460 (1996) 541, hep-th/9511030.
\bibitem{joeeric}E.~G.~Gimon and J.~Polchinski,
Phys.\ Rev.\ {\bf D54}, 1667 (1996), hep-th/9601038.


\bibitem{ocurve} K.~Dasgupta, D.~P.~Jatkar and S.~Mukhi,
Nucl.\ Phys.\  {\bf B523}, 465 (1998), hep-th/9707224;\hfill\\
B.~Craps and F.~Roose,
Phys.\ Lett.\  {\bf B445}, 150 (1998), hep-th/9808074;\hfill\\
J.~F.~Morales, C.~A.~Scrucca and M.~Serone,
Nucl.\ Phys.\  {\bf B552}, 291 (1999), hep-th/9812071;\hfill\\
B.~J.~Stefanski,
Nucl.\ Phys.\  {\bf B548}, 275 (1999), hep-th/9812088;\hfill\\
B.~Craps and F.~Roose,
Phys.\ Lett.\ {\bf B450}, 358 (1999), hep-th/9812149;\hfill\\ 
C.~A.~Scrucca and M.~Serone,
Nucl.\ Phys.\  {\bf B556}, 197 (1999), hep-th/9903145;\hfill\\
S.~Mukhi
and N.~V.~Suryanarayana,
JHEP {\bf 9909}, 017 (1999), hep-th/9907215.

\bibitem{gaugefact}P. C. Argyres, M. R. Plesser and  
A. D. Shapere, Nucl.Phys. B483 (1997) 172, hep-th/9608129.



\bibitem{ejs}N. Evans, C. V. Johnson and A. D. Shapere,
Nucl.Phys. B505 (1997) 251, hep-th/9703210.


\bibitem{wittenvarious}E. Witten, Nucl.Phys. B443 (1995) 85, hep-th/9503124. 
\bibitem{jkkm} C.~V.~Johnson, N.~Kaloper, R.~R.~Khuri and R.~C.~Myers,
Phys.\ Lett.\ {\bf B368}, 71 (1996) hep-th/9509070.

\bibitem{elevencouples}
C.~Vafa and E.~Witten,
Nucl.\ Phys.\  {\bf B447}, 261 (1995), hep-th/9505053;\hfill\\
M.~J.~Duff, J.~T.~Liu and R.~Minasian,
Nucl.\ Phys.\ {\bf B452}, 261 (1995), hep-th/9506126.




\bibitem{gibbonsmanton}G. W. Gibbons and N. S. Manton,
Nucl. Phys. {\bf B274} (1986) 183.
\bibitem{dancer}A.~S.~Dancer,
Commun.\ Math.\ Phys.\  {\bf 158}, 545 (1993).

\bibitem{zeetwo}
A.~Hanany and A.~Zaffaroni,
JHEP {\bf 9912}, 014 (1999), hep-th/9911113.

\bibitem{khoze}
N.~Dorey, V.~V.~Khoze, M.~P.~Mattis, D.~Tong and S.~Vandoren,
Nucl.\ Phys.\ {\bf B502}, 59 (1997), hep-th/9703228.

\bibitem{ahn}C.~Ahn and B.~Lee,
Phys.\ Rev.\  {\bf D59}, 026001 (1999), hep-th/9803069.


\end{thebibliography}
\end{document}